\documentclass[twocolumn,pre,showpacs,nofootinbib]{revtex4}
\usepackage{graphics,graphicx,amsfonts}

\begin{document}
\title{Incorporating Inertia Into Multi-Agent Systems}
\author{W.~C. Man}
\author{H.~F. Chau\footnote{Corresponding author: hfchau@hkusua.hku.hk}}
\affiliation{Department of Physics, University of Hong Kong, Pokfulam Road,
 Hong Kong}
\affiliation{Center of Theoretical and Computational Physics, University of
 Hong Kong, Pokfulam Road, Hong Kong}
\date{\today}

\begin{abstract} 
 We consider a model that demonstrates the crucial role of inertia and 
 stickiness in multi-agent systems, based on the Minority Game (MG).
 The inertia of an agent is introduced into the game model by allowing agents 
 to apply hypothesis testing when choosing their best strategies, 
 thereby reducing their reactivity towards changes in the environment. 
 We find by extensive numerical simulations that our game shows a remarkable
 improvement of global 
 cooperation throughout the whole phase space. In other words, the 
 maladaptation behavior due to over-reaction of agents is removed.
 These agents are also shown to be advantageous over the standard ones,
 which are sometimes too sensitive to attain a fair success rate.  
 We also calculate analytically the minimum amount of inertia needed to 
 achieve the above improvement. Our calculation is consistent with 
 the numerical simulation results. Finally, we review some related works in
 the field that show similar behaviors and compare them to our work.
\end{abstract}

\maketitle
\section{Introduction}
 There is a growing interest in studying artificial agents-interacting models 
 which are able to generate global behaviors found in social, biological  
 and economical systems~\cite{CAS}. Examples such as matching games
 ~\cite{Match} and ideal gas models of trading markets~\cite{Gas} 
 show that this approach commonly used by physicists can be nicely applied 
 to problems lay outside the discipline. One exciting fact is that these 
 artificial models, although contain simple governing rules, can still 
 generate non-trivial global cooperative behaviors~\cite{ElFarol, MG1}. 
 In these self-organized complex systems, agents can reach equilibrium states 
 through adaptation, a dynamical learning process initiated by the feedback 
 mechanism present in these systems. 

 People possesses \emph{inertia} when making decisions and switching 
 strategies in economical systems. Conceptually, this inertia is similar 
 to the one used by Newton to describe the body motions in the physical world.
 It refers to how reluctant a person is going to drop his/her current economics
 plan and look for another one, just like an
 object is reluctant to change its motion state. Inertia may originate from:
 (1)~the cost needed to change strategies,
 (2)~the low sensitivity towards a change in environment and 
 (3)~the loss-aversion behavior in human~\cite{PT} --- people loves
 to fight back from loss~\cite{CPT}. Like different bodies may have
 different mass in classical physical systems, different people may carry
 different
 inertia in economical markets. In this paper, we introduce a simple model to 
 study the idea of inertia. This model gives striking improvement of 
 cooperative behavior, such as removal of maladaptation~\cite{MichaelWong} 
 and dynamically increase of diversity among agents, without any necessity 
 to alter initial conditions and payoff mechanism.
 Actually, studies of a few variants of MG also show improvement in
 cooperations~\cite{ThermalMG,conTMG,MGSM,BatchMG,onlineMG}. We shall further
 discuss their results and compare with ours in Section~\ref{S:previous} after
 finished reporting our model and results.

 Our model is a modification of a famous econophysical model known as 
 MG, proposed by Challet and Zhang in 1997~\cite{MG1,Euro}. 
 MG is a simple game model that captures the 
 minority seeking behavior found in stock markets and resources competitions.
 (See Refs.~\cite{JohnsonBook,ChalletBook,CoolenBook} for an overview of
 econophysics and MG.)
 In MG, $N$ agents struggle to choose between two options repetitively, 
 either buy (0) or sell (1) in each turn. Those who have chosen the minority 
 sides are winners at that turn and are awarded $1$ dollar, otherwise 
 they lose $1$ dollar. The only information they received 
 is the history of the game, which is a binary bit string composed of the
 minority choices of previous $M$ turns. A strategy is a map from the set 
 of all possible histories to the set of two options. 
 If a strategy predicts the minority correctly, it is added $1$ virtual score
 point, otherwise it loses $1$ virtual score point. Each agent is assigned $S$
 strategies once and for all at the beginning of the game in order to
 aid his/her decision. 
 In standard MG, an agent makes decision based on his/her best current
 strategy at hand, namely, the one with the highest virtual score.
 
 Clearly, there are $2^{M}$ possible histories and hence $2^{2^{M}}$ 
 available strategies. However, out of the whole strategy space, 
 only $2^{M+1}$ of them are significantly different. The diversity 
 of the population is measured by $\alpha$, which is equal to $2^{M+1}/NS$.
 The smaller the $\alpha$, the more similar are the strategies hold by 
 agents. Up to first order approximation, the dynamics of MG is 
 determined by this control parameter $\alpha$.~\cite{MGAna, SavitS, SavitL}

 The most sparkling macroscopic observable in MG is perhaps the variance of 
 option attendance per agents $\sigma^{2}/N$. It represents 
 the wastage of the system and fluctuation of resources allocation; 
 the smaller $\sigma^{2}/N$, the more the whole population benefits. 
 Researchers found that $\sigma^{2}/N$ falls below the value that 
 all agents make their choices randomly in a certain range of $\alpha$. This
 indicates that agents are cooperating although they are independent 
 and selfish. More importantly, there is a phase transition at the critical 
 point $\alpha_{c}$ which divides the $\sigma^{2}/N$ against $\alpha$ curve 
 into the so-called symmetric phase ($\alpha<\alpha_{c}$) and asymmetric 
 phase ($\alpha>\alpha_{c}$).~\cite{Order} 

\section{Our Model}
 To incorporate inertia into MG, we introduce a new modification -- 
 Hypothesis Testing Minority Game (HMG). Hypothesis testing is a
 standard statistical tool to test whether 
 an effect emerged from an independent variable appears by chance or luck. 
 In the standard version of MG, the best strategy is defined as the 
 strategy with the highest virtual score. In HMG, however, an agent $k$ 
 determines his/her own best strategies by testing the following null 
 hypothesis $H_{0}$: The current strategy $\mathcal{S}_{k,0}$ 
 performs better than the other strategy $\mathcal{S}_{k,1}$ available to agent
 $k$. Note that we have restricted ourselves to the simplest case 
 $S = 2$, but the model can be easily extended to $S > 2$ cases under the same 
 formalism. This agent possesses an sustain level $I_{k} \geq 1/2$ on his/her 
 current strategy $\mathcal{S}_{k,0}$, which is the same as the confidence 
 level on the validity of the null hypothesis we commonly use in hypothesis 
 testing (that is, the acceptance area of a standard normal).  
 This $I_{k}$ defines how much he/she could sustain the under-performance of 
 $\mathcal{S}_{k,0}$ and thereby represents his/her inertia.

 The $H_{0}$ of a particular agent $k$ can be quantitatively written as 
 \begin{equation}
 H_{0} :  \frac{\Omega_{k,0}(\tau_{k})-\Omega_{k,1}(\tau_{k})}{\delta_{k}}
 > x_{k}, \label{E:Hypothesis}
 \end{equation}
 where 
 \begin{equation}
 \frac{1}{\sqrt{2\pi}}\int_{x_{k}}^{+\infty}e^{-x^{2}/2}dx = I_{k}. 
 \label{E:condition}
 \end{equation}
 Here, $\Omega_{k,j}(\tau_{k})$ is the virtual score of a particular strategy 
 $\mathcal{S}_{k,j}$ at $\tau_{k}$, where $\tau_{k}$ is the number of 
 time steps counted from his/her adoption of $\mathcal{S}_{k,0}$ for that 
 individual agent. The dominator $\delta_{k}$ represents the fluctuation 
 of strategies' performance the agent perceived. An agent $k$ would continue 
 to stick on his/her current strategy $\mathcal{S}_{k,0}$ until 
 $\Omega_{k,j}(\tau_{k})$ descends outside his/her sustain level. Then he/she
 has to admit that $H_{0}$ is not likely to be true, rejects it 
 and shift to the other strategy. After a change of strategy, the virtual 
 scores of both strategies are reset to $0$ and $\tau_{k}$ is set back to $1$.

 The higher the value of $I_{k}$, the milder his/her 
 response and the more reluctant for him/her in shifting strategies. In 
 this way, $I_{k}$ plays the role of inertia of an agent in this game. 
 Agents with $I_{k} = 1/2$ would be most similar to standard MG agent, 
 they employ strategy with the highest virtual score. However, there are still
 two differences: these HMG agents would still stick on current strategy in 
 case of a tie in virtual scores, and the virtual scores will be reseted after 
 shifts in strategies.

 We remark that randomness are involved in only three places in HMG, namely,
 (1)~the initial assignment of strategies and inertia;
 (2)~the choice of a new strategy in case of a tie in the virtual scores of the
 alternative strategies when a player has decided to drop the current one;
 as well as
 (3)~the determination of the winning side in case of a tie. Thus, the dynamics
 of HMG is deterministic when played by an odd number of agents each carrying
 $2$ strategies.

 \section{Pure Population With Random Walk Approximation}
 We have performed extensive numerical simulation on our model.
 With the presence of inertia among agents, every agent needs a longer time 
 to make up his/her mind and the equilibration time in HMG is lengthened. 
 We take the value of $\sigma^{2}$ every 1,000 time 
 steps and regard the system as having equilibrated if the percentage
 difference of successive measurement is smaller than $\epsilon = 10^{-6}$. 
 Upon equilibration, we take our measurement by recording the dynamics of the
 next 25,000 time steps. Furthermore, we repeat this data taking procedure
 150 times, each with an independent set of initial conditions.

 In a population where everyone tries to cling on the minority side as long 
 as possible, agents may have different inertia $I_{k}$, and some may
 have no inertia at all (standard MG agents).
 We first study the behavior of HMG when every agent has the same $I_{k}$.
 (We shall move on to study the mixed population case in later sections.)
 We begin our study by determining the value of
 $\delta_{k}$, a perception of agents on the fluctuation of virtual
 score difference between two strategies. A naive guess would be assuming 
 $\Omega_{k,j}(\tau_{k})$ performs random walk for all strategies $j$ 
 throughout the game, then $\delta_{k}$ equals $\sqrt{2 \tau_{k}}$.  

 \begin{figure}[ht]
 \centering
 \includegraphics*[scale=0.3]{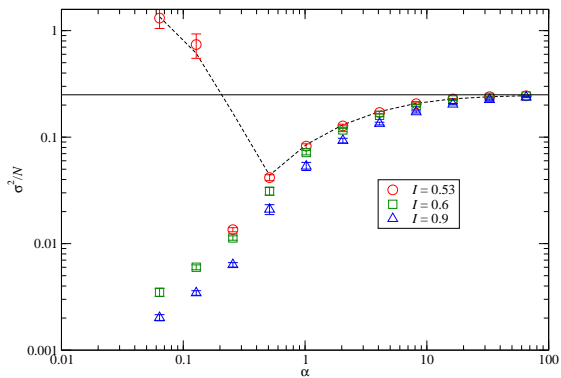}
 \caption{The variance of attendance per agent $\sigma^{2}/N$ against the 
 system complexity $\alpha$ for $I$ equals (a) 0.53, (b) 0.6 and (c) 0.9, 
 setting $\delta_{k}$ equals $\sqrt{2\tau_{k}}$. Here, $N$ = 501 and $S$ = 2. 
 The dashed line represents the $\sigma^{2}/N$ curve in standard MG.}
 \label{F:TMG}
\end{figure}

 \begin{figure}[ht]
 \centering
 \includegraphics*[scale=0.3]{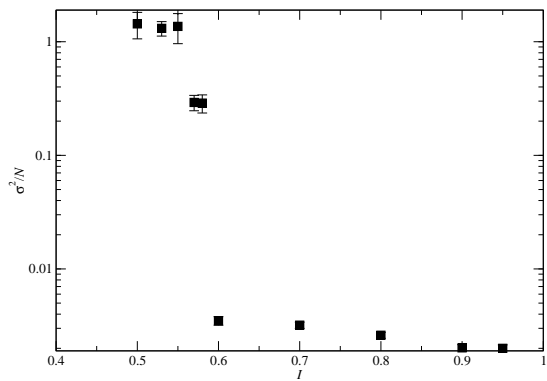}
 \caption{The variance of attendance per agent $\sigma^{2}/N$ against the 
 $I$ at $\alpha$ = 0.06, setting $\delta_{k}$ equals $\sqrt{2\tau_{k}}$. 
 Here, $N$ = 501, $S$ = 2 and $M$ = 5.}
 \label{F:TMG_ac}
\end{figure}

 Fig.~\ref{F:TMG} shows a plot of the variance of attendance for a 
 particular option $\sigma^{2}/N$ against the control parameter 
 $\alpha$ for different 
 inertia $I$, with $\delta_{k}$ set to $\sqrt{2\tau_{k}}$.
 There is a huge drop of $\sigma^{2}/N$ when $I$ is sufficiently large, 
 especially in symmetric phase when $\alpha$ is small (see
 Fig.~\ref{F:TMG_ac}).
 Not just the maladaptation in symmetric phase is greatly reduced, but 
 the cooperation between agents is also improved in the asymmetric phase 
 for certain values of $I$.

\begin{figure}[ht]
 \centering
 \includegraphics*[scale=0.3]{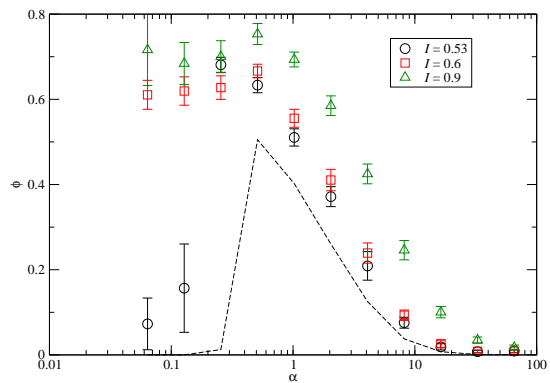}
 \caption{The frozen probability $\phi$ against $\alpha$ for different  
 $I$ by setting $\delta_{k}$ equals $\sqrt{2\tau_{k}}$. The dashed line
 represents the frozen probability of the standard agents in MG. 
 Here, $N$ = 501, $S$ = 2 and $M$ = 5.}
 \label{F:TMG_frozen}
\end{figure}

 The reduction of system wastage in the asymmetric phase ($\alpha >
 \alpha_{c}$) is believed to be resulted by increasing stickiness of agents on 
 current strategies and elongating their observing time. This leads to 
 an increase of frozen agents (see Fig.~\ref{F:TMG_frozen}) and an more 
 effective crowd-anticrowd cancellation, succeeding in better cooperation.
 \cite{Order,CAC1, CAC2}
 
\begin{figure}[ht]
\centering
\includegraphics*[scale=0.3]{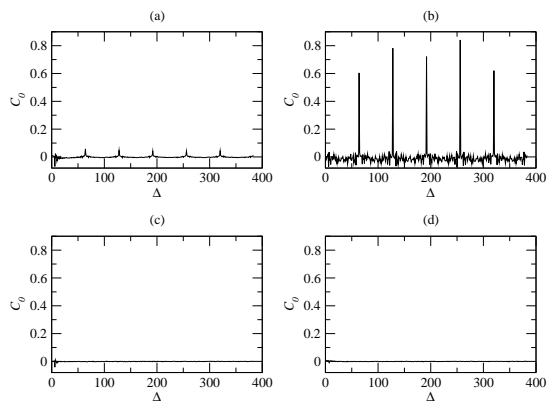}
 \caption{The autocorrelation of attendance $C_{0}$ against various 
 interval $\Delta$ on cases (a) standard MG, (b) $I$ =  0.53, (c) $I$ = 0.6
 and (d) $I$ = 0.9 averaged over 50 independent runs. Here, $N$ = 501, 
 $M$ = 5 and $S$ = 2. }
 \label{F:TMG_corr}
\end{figure}

 However, things become more complicated when $\alpha < \alpha_{c}$.
 From now on, this article will focus on the striking improvement of 
 cooperation in the symmetric phase. The removal of maladaptation in 
 this region is directly related to the disappearance of periodic dynamics 
 that normally present in the standard MG. The periodic dynamics 
 is a result of oversampling of strategy space and common 
 zero initial conditions among agents when $\alpha < \alpha_{c}$, 
 accounting for the high volatility in the symmetric phase. 
 It is reflected in a prominent period $2^{M+1}$ peak in the autocorrelation 
 of the attendance time series of a particular option
 \cite{MichaelWong, Mem, Lee, SavitL, SavitS}. Fig.~\ref{F:TMG_corr} 
 shows an evidence of this postulate: as shown from the autocorrelation 
 function, periodic dynamics appears in the case $I$ = 0.53 which has 
 high $\sigma^{2}/N$ in Figs.~\ref{F:TMG} and~\ref{F:TMG_ac}, while the
 low $\sigma^{2}/N$ cases $I$ = 0.6 and $I$ = 0.9 show no trace of this signal.

 What is the critical limit of $I$ in order to remove the maladaptation? 
 To answer this, we have to look closely into the periodic dynamics 
 that governs the maladaptation in the symmetric phase.
 Earlier study stated that virtual scores of strategies are likely 
 to reset to $0$ every $2^{M+1}$ number of steps through the periodic 
 dynamics in the symmetric phase. Initially all strategies have $0$ score
 point, whenever a strategy $\beta$ wins a bet in a particular $\mu$, 
 most agents would rush to $\beta$ which is $2$ score points ahead its 
 anti-correlated partner $\bar{\beta}$ in the next appearance of $\mu$.  
 It is likely that they would lose due to this overcrowding. In this manner, 
 the virtual scores of all strategies are reset at this stage. 
 This loop repeats with interval $2^{M+1}$ and leads to 
 the large fluctuation of option attendance in the symmetric
 phase.~\cite{Symmetric}
 
 Therefore, the question becomes when this reset and oscillate mechanism 
 will disappear. Actually, the periodic dynamics is destroyed when 
 agents are no longer sensitive enough to immediately shift to a 
 strategy standing out after winning a bet. The criteria for this 
 situation to occur is given by:
 \begin{equation}
  \frac{-2}{\sqrt{2 \cdot 2^{M+1}}} < x_k,
  \label{TMG_limit}
 \end{equation}
 where $x_{k}$ satisfies Eq.~(\ref{E:condition}).
 If the value of $I$ satisfies the inequality~(\ref{TMG_limit}), 
 agents would no longer be constrained by the periodic dynamics every $2^{M+1}$ 
 steps.  Then, a re-recognizing process will draw in. In the standard MG, all 
 identical strategies have same virtual scores throughout the game. However, 
 in HMG agents would clear all virtual scores after changing strategies. 
 This move is done in multifarious time steps for different agents, depending
 on the combination of strategies in their hands. Hence, the scores of 
 identical strategies eventually diverges if they are hold by different agents,
 and these strategies may be employed again in multifarious time in the future.
 The net effect of this re-recognizing process is diversifying  
 strategies in the population intrinsically. In this way, 
 both the oversampling and overcrowding found in the symmetric phase 
 are relaxed, lowering the volatility. 

 For instance, when $M$ = 5, the limit $x_{c}$ equals $-2/\sqrt{2 \cdot
 2^{5+1}} = -0.177$; that is, $I \simeq 0.57$. This criteria is confirmed
 in Figs.~\ref{F:TMG_ac} and~\ref{F:TMG_corr} --- all cases that show no 
 periodic dynamics satisfies Eq.~(\ref{TMG_limit}) and have low variances. 
 Note that for the cases where $I$ does not exceed this limit, their 
 correlation signals are much stronger than that of the standard MG 
 (see Fig.~\ref{F:TMG_corr}b). It is because the dynamics of HMG is more
 deterministic than that of the standard MG as HMG agents will
 continue to stick on current strategy when facing
 a tie in strategy virtual scores, which happens during a reset. That means 
 their actions repeat during this reset and the system path is more likely 
 to repeat, resulting in stronger correlation. This is like removing the 
 random dice in standard MG when facing a tie in virtual scores, a 
 periodic signal as strong as this case is also obtained. 

 \section{Pure Population With Runtime $\delta_{k}$}  
 Actually, the movement of the virtual score difference between two strategies 
 is not likely to perform random walk. Another possible way in perceiving
 $\delta_{k}$ is to put it as the actual standard deviation of 
 this difference in runtime, which represents a more realistic market 
 scenario. That is, 
 \begin{equation}
 \delta_{k} = \sqrt{\langle
 (\Omega_{k,0}(\tau_{k})-\Omega_{k,1}(\tau_{k}))^2 \rangle_{\tau_k} - \langle
 \Omega_{k,0}(\tau_{k})-\Omega_{k,1}(\tau_{k}) } \rangle_{\tau_{k}}^2.
 \label{E:run_time}
 \end{equation} 

\begin{figure}[ht]
 \centering
 \includegraphics*[scale=0.3]{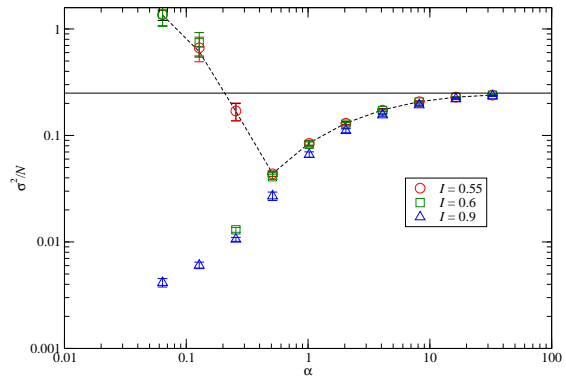}
 \caption{The variance of attendance per agent $\sigma^{2}/N$ against the 
 system complexity $\alpha$ for $I$ equals (a) 0.55, (b) 0.6 and (c) 0.9, 
 with $\delta_{k}$ given by Eq.~(\ref{E:run_time}). Here, $N$ = 501 and 
 $S$ = 2. The dashed line represents the $\sigma^{2}/N$ curve in standard MG.}
 \label{F:VarHMG}
 \end{figure}

\begin{figure}[ht]
 \centering
 \includegraphics*[scale=0.3]{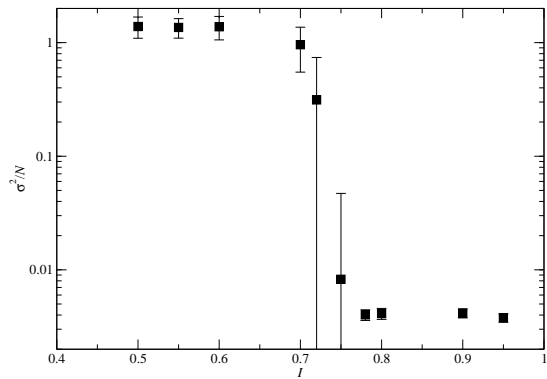}
 \caption{The variance of attendance per agent $\sigma^{2}/N$ against 
 $I$ at $\alpha$ = 0.06, setting $\delta_{k}$ satisfying 
 Eq.~(\ref{E:run_time}).  Here, $N$ = 501, $S$ = 2 and $M$ = 5.}
 \label{F:VarHMG_ac}
\end{figure}

\begin{figure}[ht]
\centering
\includegraphics*[scale=0.3]{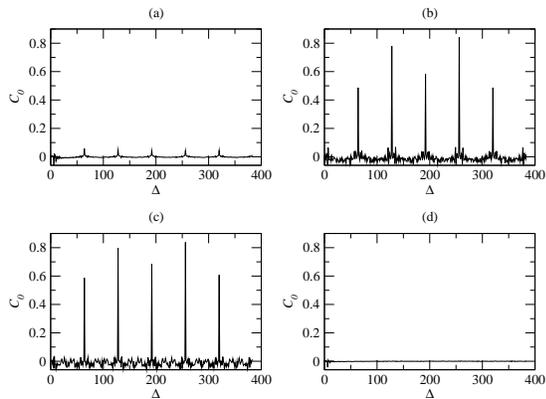}
 \caption{The autocorrelation of attendance $C_{0}$ against various 
 interval $i$ on cases (a) standard MG, (b) $I$ =  0.55, (c) $I$ = 0.6 
 and (d) $I$ = 0.9 averaged over 50 independent runs. Here, $N$ = 501, 
 $M$ = 5 and $S$ = 2 and $\delta_{k}$ given by Eq.~(\ref{E:run_time}). } 
 \label{F:VarHMG_corr}
\end{figure}

 The results are very similar to the previous case, which are shown 
 in Figs.~\ref{F:VarHMG}--\ref{F:VarHMG_corr}. 
 However, the critical value of $I$ for the system to escape 
 from the grip of periodic dynamics appears to be higher. Remind that 
 the virtual score difference of two strategies performs random walk 
 with following step sizes and probabilities $p$:
 \begin{equation}
 \Omega_{k,0} - \Omega_{k,1} = \left\{ \begin{array}{ll}
					+2 & \mbox{ with $p$ = $1/4$},\\
					-2 & \mbox{ with $p$ = $1/4$},\\
					0 & \mbox{ with $p$ = $1/2$}.
					\end{array}
				\right.
 \end{equation} 
 Meanwhile, the presence of periodic dynamics ensure a reset every $2^{M+1}$ 
 number of time steps. We can approximately calculate the average variance 
 by considering all possible traveling paths, which equals $2^{M+1}/12$ 
 (detail mathematics is shown in the Appendix). For instance, when $M$ = 5, 
 the critical value for the periodic dynamics to disappear is $x_{c} < 
 -2/\sqrt{2^{5+1}/12} = -0.866$; that is, $I \simeq 0.81$. This value of $I$
 is consistent with the data presented in Figs.~\ref{F:VarHMG_ac}
 and~\ref{F:VarHMG_corr}. Again, we believe that after the breaking of periodic
 dynamics, the re-recognizing process mentioned in the previous section 
 comes in and diversifies the strategy space, resulting in a drop of 
 fluctuation.

 \section{Mixed Population With Standard MG Agents}
 It is already clear that a pure population of agents having inertia reduces 
 system wastage. Now it is instructive to study whether these agents 
 (sticky agents) is advantageous over standard MG agents (sensitive agents) 
 in a mixed population. 

 \begin{figure}[ht]
 \centering
 \includegraphics*[scale=0.3]{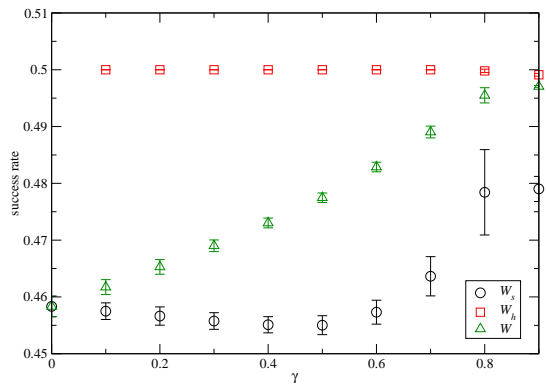}
 \caption{A plot of success rate of sticky agents $W_{h}$, sensitive 
 agents $W_{s}$ and the whole population $W$ against the fraction of 
 sticky agents $\gamma$ in the mixed race. There are total $501$ 
 agents, with $M$ = 5, $S = 2$ and $I$ = 0.9.}   
 \label{F:Mix}
\end{figure}
  
\begin{figure}[ht]
 \centering
 \includegraphics*[scale=0.35]{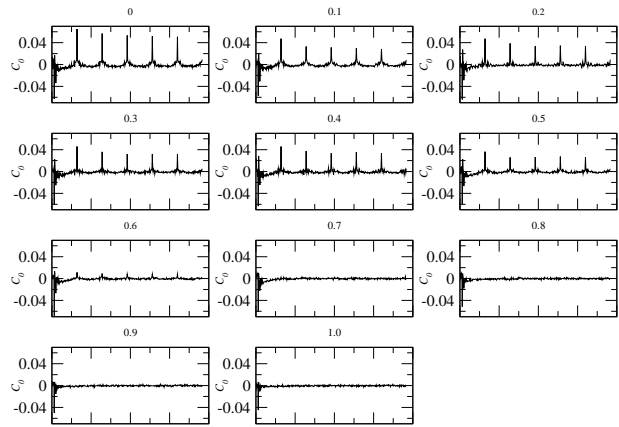}
 \caption{The autocorrelation of attendance $C_{0}$ for different $\gamma$ 
 averaged over 50 independent runs. Here, $N$ = 501, $M$ = 5, $S$ = 2
 and $I$ = 0.9.}
 \label{F:Mix_corr}
\end{figure}

 Fig.~\ref{F:Mix} gives the success rates of both races against 
 $\gamma$ in the mixed population with $I$ = 0.9, 
 where $\gamma$ is the fraction of sticky agents in the population. 
 Clearly, these sticky agents take advantages of the sensitive agents for 
 whole range of $\gamma$, they successes in maintaining their success rates 
 close to $0.5$. The sensitive agents are believed to be tightened by 
 the periodic dynamics, making them to keep on losing. On the other hand, 
 sticky agents are likely to win more frequently as they are resistant 
 to follow the oscillation. Note that the whole population also benefits 
 from adding in more sticky agents (see the triangles in Fig.~\ref{F:Mix}).
 When $\gamma$ is increased up to about $0.6$, $W_{s}$ starts to rise. 
 It is because the crowd of sensitive agents is no longer large enough to 
 override the net actions made by sticky agents, and therefore 
 there is no more periodic dynamics existing. Fig.~\ref{F:Mix_corr} confirms
 our suspicion, the periodic dynamics disappear around $\gamma$ = 0.6.
 We have also performed simulations on mixed population of sensitive agents 
 and sticky agents with other values of $I$. As expected, sticky agents
 are only advantageous with $I$ exceeds the critical value that 
 allow them to escape from periodic dynamics mentioned in the last section. 
 Otherwise, all agents in the whole population would still suffers from 
 overcrowding and no one will be benefited.
  
\section{Previous studies in the literature and comparison with our results}
\label{S:previous}
 The reduction of the volatility by modifying the rules or the initial
 conditions of the standard MG is not a new idea in the field, 
 especially for the symmetric phase. A few previous studies have shown 
 results quite similar to that of the HMG.
 Here, we would like to first give a short review of these works and to
 compare them with our study.

\subsection{Thermal Minority Game}
 Cavagna \emph{et al.} proposed the Thermal Minority Game
 (TMG)~\cite{ThermalMG}, which
 adds stochasticity into the standard MG. In TMG, an agent does not
 employ the strategy with highest virtual score strict a way, rather 
 he/she would use a strategy with probability calculated according to its
 virtual score and a fixed inverse temperature $T$. In other words, the 
 employment of strategies by agents become probabilistic with the degree 
 of stochasticity depending on $T$. They found that for certain range of $T$, 
 the volatility in the game is reduced in most range of the control 
 parameter $\alpha$. That is, in both symmetric phase and asymmetric phases,
 TMG succeeds in raising the degree of cooperation between agents by 
 introducing noises into the decision process of strategy selection for
 individual agents.

 In search of the continuous time dynamics of TMG, Garrahan \emph{et al.}
 confirm by numerical simulation that the dynamics in the symmetric phase 
 of MG is sensitive to initial conditions. In particular, they reported 
 that the volatility would drop far from the original value if
 random initial conditions to strategies (with $O(1)$ initial virtual
 scores for a population of 100 agents) are assigned at the beginning of the
 game.~\cite{conTMG}

\subsection{Nash equilibrium}
 In searching the replica solution and the Nash equilibrium for the symmetric
 phase of the standard MG, Challet \emph{et al.} found that the
 Nash equilibrium 
 is not unique and agents at these equilibria use pure strategies (that is, 
 they either always choose $1$ or always choose $0$).~\cite{MGSM} In Nash 
 equilibrium, agents perform much better than in the standard MG, 
 the volatility is greatly suppressed in the symmetric phase.

\subsection{Consideration of agents' own market impact in evaluation of 
 strategy}
 Challet \emph{et al.} try to let the agents consider their own impact on the 
 market during the evaluations of all strategies available to them. That is, 
 the virtual score of a strategy is proportional to the cumulated payoff 
 the agent would have received had he or she always played the strategy. 
 Although the difference between this evaluation of virtual score and the 
 original one is believed to be small ($\sim 1/\sqrt{N}$), the volatility is 
 found to be far lower than the original MG. This difference is not negligible 
 because of finite size effect and the high degree of over-sampling of the 
 strategy space when $\alpha < \alpha_{c}$. However, this setting is 
 computational intensive and unrealistic, as people in real market
 usually can only obtain information on his/her own current wealth and unlikely
 to try out all strategies.~\cite{MGSM}

\subsection{The analytical solutions of batch minority game and the on-line 
 minority game}
 In batch minority game, the virtual score of a particular strategy is updated 
 as discrete accumulated effect of order $N$ iterations in the standard 
 MG model, whereas the MG having the original updating method can be viewed 
 as a ``online'' minority game in the neural network sense. After adding in
 stochasticity, initial evaluations and generalizing these game to continuous 
 time limit,  Coolen's group has extensively written out the analytical 
 solutions of these two versions of MG. They found that in symmetric phase
 their theory pointed at the existence of a critical value for the initial
 strategy valuations above the system would revert to a state with vanishing
 volatility.~\cite{BatchMG, onlineMG}

\subsection{Introduction of diversity}
 Wong \emph{et al.} pointed out in~\cite{MichaelWong} that the maladaptation 
 observed in the symmetric phase in the standard MG is originated 
 from the fact that initial virtual scores of all strategies are the same.
 They then studied the effect of introducing diversity $R/N$
 into the game, where $R$ is the range of randomly assigned initial 
 scores to strategies at the beginning of the game and $N$ is the 
 number of agents. They found that by increasing the diversity, 
 the maladaptive behaviour observed in the symmetric phase $\alpha 
 < \alpha_{c}$ is reduced and hence the cooperation among agents 
 is promoted.

\subsection{Comparison to our model}
 From the above studies, we can conclude that the volatility would suppressed 
 under following conditions: (1)~randomly allocating initial strategy score
 over a critical value, (2)~adding in noise or stochasticity in choosing a
 strategy, (3)~assigning pure strategy or (4)~taking an agent's impact on
 market into account when evaluating all strategies. 

 Firstly, we would like to stress that the main focus in this article 
 is to provide a simple formalism to incorporate inertia into a multi-agent 
 system such as MG, as well as recording its influence to the dynamics 
 of the game. In HMG, there is no prior preference in strategies for they
 have the same initial virtual score. Unlike the standard MG, soon after
 the commencement of HMG, the preference of a strategy is determined
 by both the virtual score differences between strategies at hand and
 inertia $I_k$ of agent $k$. Through the presence of inertia, each agent will
 gradually develop their own preference in strategies through dynamical
 adaptation. In this respect, even though the presence of inertia may 
 eventually lead to difference views of an identical strategy between agents, 
 this is achieved by an adaptive process through the dynamics of the system 
 but not by artificially assigning a spread of initial virtual scores. This
 is a marked difference between HMG and the works of Wong
 \emph{et al.}~\cite{MichaelWong},
 Garrahan \emph{et al.}~\cite{conTMG} as well as
 Coolen \emph{et al.}~\cite{BatchMG,onlineMG}.
 More importantly, the spreading of initial virtual scores of strategies would 
 only leads to a drop of volatility in the symmetric phase, but not 
 the asymmetric phase. In HMG, however, there is a global improvement in 
 both phases for certain value of $I$. 

 We believed the TMG presents results most similar to our game. In both 
 case, the degree of cooperation are raised in most range of $\alpha$. 
 However, as mentioned previously, TMG achieve this by adding stochasticity 
 and noise into agents' choice of best strategies. Meanwhile, 
 in HMG agents are deterministic when choosing their best strategies:
 they stick to their current strategy until it is outperform to certain 
 threshold, this does not involve any stochasticity. In fact, the dynamics of
 HMG is deterministic when played by an odd number of agents each carrying $2$
 strategies.

 Lastly, we think that using pure strategies and taking agents' themselves 
 into account when evaluating all their strategies are impractical and 
 unrealistic situations. Our model provide a natural, realistic way to 
 prompt cooperation, meanwhile demonstrating the effect of stickiness 
 when people moving around investment strategies.

 \section{Conclusions}
 We have successfully introduced the concept of inertia into the Minority Game,
 which shows a remarkable improvement of cooperation among agents in 
 most range of $\alpha$, especially in the symmetric phase 
 $\alpha < \alpha_c$. We also compare our findings with a few variants of MG
 reported in the literature. We calculated the critical values of inertia 
 needed to uplift the cooperation behaviors, which depends on how agents 
 perceive the fluctuation of virtual score difference between strategies. 
 This reduction of sensitivity among agents is found to be useful in 
 removing maladaptation due to over-reaction. In contrast, 
 if every action is smooth and all agents response to information in no time, 
 they will suffer from a overcrowd loss easily. Meanwhile, agents carrying 
 stickiness seems to perform much better than sensitive agents. 
 Our findings suggest that inertia (or stickiness) is crucial and beneficial 
 to a society. It is hoped that the role of inertia will be 
 investigated in more detail based on our model HMG, such as the effect of 
 giving a diversifying range of inertia to a population.
 It is also instructive to apply our method of modeling inertia to study 
 inertia effect in other multi-agent systems.
 
\acknowledgments
 We thank the Computer Center of HKU for their helpful support in providing 
 the use of the HPCPOWER System for simulations reported in this article. 
 K. M. Lee, F. K. Chow, K. H. Ho and C. C. Leung are gratefully acknowledged 
 for their fruitful discussions.

\appendix
\section*{Appendix} 
\begin{figure}[ht]
\centering
\includegraphics*[scale=0.3]{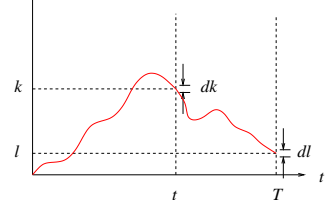}
 \caption{A schematic sketch showing a typical random walk of particle travels 
 for $T$ time step.} 
 \label{F:path}
\end{figure}

In this appendix, we consider a simple random walk of a cumulative sum $x_t$
after time $t$. At each step, $x_t$ increases (decreases) by $1$ if moves 
upward (downward) with probability $1/2$. We also impose a 
boundary condition that the sum is equal to $0$ at both $t = 0$ and $t = T$. 
A schematic diagram is shown in Fig.~\ref{F:path}.
Under such constraint, we find that the variance of $x_t$ averages over all 
possible paths $\sigma_r^2 = \frac{t(T-t)}{4T}$. Using this formula, we can 
evaluate the average standard deviation of virtual score difference of an 
agent's strategies within a period $2^{M+1}$, that is the $\delta_{k}$ 
mentioned in Eq.~(\ref{E:run_time}).

 First, we need to know the probability of $x_t$ within $k$ and $k+dk$,
 which is given by~\cite{Springer}
\begin{equation} \label{E:prob}
  P(k \leq x_t \leq k+dk)    \approx   \sqrt{\frac{2}{\pi t }} e^{-2k^{2}/t}dk.
\end{equation}
Hence, the probability of the cumulative sum $x_t$ to be within
$k$ and $k+dk$ at time $t$ and $x_T$ to be within $l$ and $l+dl$ 
at time $T$ can be expressed by
\begin{eqnarray}
& & P(k \leq x_{t} \leq k +dk \,\, {\rm and}  \,\,l \leq x_{T} \leq l+dl) 
\nonumber  \\
& = & P(k \leq x_{t} \leq k+dk) \cdot P(l-k \leq x_{T-t} \leq l-k +dl ) 
\nonumber \\
\label{E:prob_and}
\end{eqnarray}
 where $t \leq T$. The equality follows from the fact that the discrete steps 
 size is equal to $1$. Using the Eqs.~(\ref{E:prob}) and~(\ref{E:prob_and}),
 the conditional probability
\begin{eqnarray}
& &   P(k \leq x_{t} \leq k+dk | l \leq x_{T} \leq l+dl) \nonumber \\
  & = & \frac{P(k \leq x_{t} \leq k +dk \,\, {\rm and} \,\,l \leq x_{T} \leq
 l+dl)}{P( l\leq x_T \leq l +dl)}\\
  & = & \frac{\sqrt{\frac{2}{\pi t}}e^{-2k^{2}/t} \cdot
 \sqrt{\frac{2}{\pi (T-t)}}e^{-2(l-k)^{2}/(T-t)}dk  dl}{ \sqrt{\frac{2}{\pi
 T}}e^{-2l^{2}/T}dl}.
\end{eqnarray}
By the boundary condition, $0 \leq x_T \leq dl$, then we have
\begin{eqnarray}
& & P(k \leq x_{t} \leq k+dk | 0 \leq x_T \leq dl) \nonumber \\
&=& \sqrt{\frac{2T}{\pi t(T-t)}} e^{-2T k^2 /t(T-t)}dk.
\end{eqnarray}
Therefore, the variance $\sigma_r^2$ averaged over all possible paths: 
\begin{eqnarray}
\sigma_r^2 &=& \int_{-\infty}^{+\infty}k^{2} P(k \leq x_{t} 
\leq k+dk | 0 \leq x_T \leq dl) \\
& = & \frac{t(T-t)}{4T}.
\end{eqnarray}

In order to calculate $\delta^{2}_{k}$, we should rescale $\sigma_r^2$ because
the virtual score difference of an agent's strategies can move two steps 
upward ($+2$), two steps downward($-2$) or keep stationary($0$). Hence, by 
approximating the travel time $T$ consists infinity number of time steps: 
 \begin{eqnarray}
 \delta^{2}_{k} & = & 2 \cdot \frac{1}{T}\int_{0}^{T}\frac{t(T-t)}{4T}dt =
 \frac{T}{12}.
 \end{eqnarray}
 where $\delta_{k}$ is the perceived fluctuation mentioned in 
 Eq.~(\ref{E:run_time}).

\end{document}